\documentclass[10pt,twocolumn,letterpaper]{article}

\usepackage{graphicx}
\usepackage{amsmath}
\usepackage{amssymb}
\usepackage{booktabs}
\usepackage{placeins}  
\usepackage{pifont} 
\usepackage{multirow}
\usepackage{graphicx}
\usepackage{hyperref}
\usepackage{float}  

\usepackage{hyperref}  

\usepackage[capitalize]{cleveref}
\crefname{section}{Sec.}{Secs.}
\Crefname{section}{Section}{Sections}
\Crefname{table}{Table}{Tables}
\crefname{table}{Tab.}{Tabs.}

\def\cvprPaperID{*****} 
\def\confName{CVPR}
\def\confYear{2022}

\begin{document}
\title{MedBLIP: Fine-tuning BLIP for Medical Image Captioning}

\author{Manshi Limbu\\
George Mason University\\
{\tt\small klimbu2@gmu.edu}
\and
Diwita Banerjee\\
George Mason University\\
{\tt\small dbanerj@gmu.edu}
}
\date{May 10, 2025}
\maketitle

\begin{abstract}
Medical image captioning is a challenging task that requires generating clinically accurate and semantically meaningful descriptions of radiology images. While recent vision-language models (VLMs) such as BLIP, BLIP2, Gemini and ViT-GPT2 show strong performance on natural image datasets, they often produce generic or imprecise captions when applied to specialized medical domains. In this project, we explore the effectiveness of fine-tuning the BLIP model on the ROCO dataset for improved radiology captioning. We compare the fine-tuned BLIP against its zero-shot version, BLIP-2 base, BLIP-2 Instruct and a ViT-GPT2 transformer baseline. Our results demonstrate that domain-specific fine-tuning on BLIP significantly improves performance across both quantitative and qualitative evaluation metrics. We also visualize decoder cross-attention maps to assess interpretability and conduct an ablation study to evaluate the contributions of encoder-only and decoder-only fine-tuning. Our findings highlight the importance of targeted adaptation for medical applications and suggest that decoder-only fine-tuning (encoder-frozen) offers a strong performance baseline with 5\% lower training time than full fine-tuning, while full model fine-tuning still yields the best results overall. \href{https://github.com/Diwita19/Medical_Image_Captioning}{\texttt{github.com/Med\_Img\_Captioning}}

\end{abstract}

\section{Introduction}

Automatic image captioning has become a key task in VLMs, enabling systems to describe visual content in natural language. While state-of-the-art models such as BLIP \cite{li2022blip}, CLIP \cite{radford2021learning}, BLIP-2\cite{li2023blip2bootstrappinglanguageimagepretraining}, Gemini 1.5 Flash and ViT-GPT2 \cite{vit_gpt2_captioning} achieve strong performance on general datasets like MS-COCO, Flickr30k and others, they often struggle when applied directly to specialized domains like medical imaging. Radiology images differ significantly from natural images in both content and visual structure, and their corresponding captions often include technical language, anatomical references, and clinical findings that general models are not trained to handle.

Medical image captioning holds promise for a range of applications, including automated documentation, assistive diagnostics, and structured report generation. However, most pretrained models tend to generate overly generic or factually incorrect descriptions when exposed to radiology data. This presents a critical challenge, especially in high-stakes clinical environments where precision and factual grounding are essential.

In this project, we investigate whether fine-tuning a state-of-the-art VLM BLIP on the ROCO dataset can improve caption relevance and accuracy for radiology images. We compare the zero-shot BLIP model against its fine-tuned counterpart, and other baseline models; BLIP-2, BLIP-2 Instruct, Gemini 1.5 Flash and ViT-GPT2. Additionally, we explore attention visualization to understand how model predictions relate to image regions, and conduct an ablation study to assess the impact of fine-tuning different components of the model.

\textbf{Our contributions are as follows:}
\begin{itemize}
    \item We fine-tune the BLIP model on radiology images and demonstrate improved performance across multiple evaluation metrics.
    \item We perform a comparative analysis between BLIP (base and fine-tuned) with other transformer based architectures; BLIP-2, BLIP-2 Instruct, Gemini 1.5 Flash and ViT-GPT2.
    \item We visualize decoder cross-attention maps to interpret the model’s token-to-image alignment.
    \item We conduct an ablation study to assess the impact of fine-tuning the encoder-only, decoder-only, as well as both.
\end{itemize}

\section{Related Work}
Image captioning has evolved from traditional recurrent models to large-scale transformer-based architectures that leverage multimodal pretraining. This progression is especially relevant in the medical domain, where models must not only generate grammatically correct text but also convey accurate clinical semantics.

\subsection{ Classical Approaches: CNN-LSTM}
Early captioning models \cite{vinyals2015show, xu2015show} employed a convolutional neural network (CNN) for image feature extraction followed by a long short-term memory (LSTM) network for sequential text generation. While these approaches laid foundational work in visual-language modeling, they were limited in capturing long-range dependencies and lacked contextual grounding beyond localized attention.

\subsection{ Transformer-Based Architectures} 
With the advent of transformers, models like ViT-GPT2 \cite{vit_gpt2_captioning}, BLIP-2, BLIP-2 Instruct, and Gemini 1.5 Flash have replaced traditional recurrent components with a Vision Transformer (ViT) encoder and a GPT-style autoregressive decoder. These models, trained on large-scale datasets such as MS-COCO, Flickr30k, and LAION, have demonstrated strong performance in general image captioning. However, their effectiveness often deteriorates in specialized domains such as radiology due to mismatches in vocabulary, style, and image distribution.

The BLIP (Bootstrapped Language-Image Pretraining) framework \cite{li2022blip} marked a significant step forward by unifying contrastive learning, image-text matching, and generative objectives within a single architecture. BLIP supports both discriminative and generative tasks and has shown strong performance in zero-shot and few-shot settings. Recent adaptations for medical imaging \cite{chen2023medblip, blip2_medical} further demonstrate the framework’s potential for radiology-focused applications.

\subsection{Medical Captioning Datasets and Domain Challenges}
The \textbf{ROCO dataset} \cite{roco2018}, along with its newer release \textbf{ROCOv2} \cite{rocov2}, provides radiology images paired with figure captions extracted from open-access publications. Other commonly used datasets in this space include \textbf{IU Chest X-ray} \cite{iu_xray} and \textbf{MIMIC-CXR}, though the latter focuses more on radiology report generation than descriptive image captioning.

Despite the availability of these datasets, radiology captioning presents domain-specific challenges \cite{liu2023systematic}, including hallucinated findings, ambiguous anatomical references, and limited data for rare modalities. Recent works such as \textbf{MedBLIP} \cite{chen2023medblip}, \textbf{LLaVA-Med}, and \textbf{Visual Cluster Grounding} \cite{zhang2022visual} aim to improve vision-language alignment in the medical setting. Additionally, the \textbf{ImageCLEF Medical Caption} competition \cite{imageclef2024, vannguyen2024uitdarkcowteamimageclefmedicalcaption, ImageCLEFmedicalCaptionOverview2024} reflects growing interest in benchmarking and evaluating domain-adapted captioning systems.

\subsection{Evaluation and Interpretability}
 
Recent works emphasize the importance of both standardized captioning metrics—such as SPICE, CIDEr, BERTScore (both original BERT and Bio\_ClinicalBERT), and cosine similarity with Bio\_ClinicalBERT.  Additionally, \textbf{attention-based visualizations} have emerged as a key tool for model interpretability. These evaluation strategies are particularly valuable in the medical domain, where factual correctness and semantic alignment take precedence over surface-level fluency. However, as noted in the systematic review by Liu et al. \cite{liu2023systematic}, most deep learning models for medical image captioning continue to face challenges in generalization and interpretability. To address this, we complement traditional metrics with a \textbf{clinical correctness evaluation table}, which assesses captions based on modality, laterality, anatomical specificity, and diagnostic accuracy which further helps identify if the model hallucinate findings or misidentify clinical structures for safety-critical applications like radiology.

\section{Methodology}
\subsection{Dataset Preparation}
We use the ROCO dataset \cite{roco2018,rocov2}, which consists of radiology images paired with figure captions extracted from open-access biomedical publications. We focus on a subset of images relevant to common modalities such as chest X-rays, MRIs, CT scans, and ultrasounds. Each image is paired with a short natural language description. All images were resized to 384×384 to match the input requirements of BLIP and other transformer-based models. Captions were tokenized using the associated model tokenizer, and the dataset was formatted to Hugging Face’s `datasets` structure for compatibility.

\subsection{Model Architectures}
We experiment with the following models:
\begin{itemize}
    \item \textbf{ViT-GPT2}: A transformer-based encoder-decoder model built using the `VisionEncoderDecoderModel` class with ViT as the encoder and GPT-2 as the decoder \cite{vit_gpt2_captioning}.
   \item \textbf{BLIP (Base)}: The pretrained BLIP image-captioning model \cite{li2022blip}, used in zero-shot mode without task-specific fine-tuning.
   \item \textbf{BLIP (Fine-tuned)}: The BLIP model with a ViT-B/16 encoder and transformer decoder is fine-tuned on the ROCO dataset using cross-entropy loss with teacher forcing, gradient accumulation, mixed-precision, and a linear learning rate schedule.
   \item \textbf{BLIP-2}: A vision-language model combining a ViT encoder, Q-Former adapter, and frozen language decoder (e.g., Flan-T5 or OPT), used in zero-shot mode \cite{li2023blip2bootstrappinglanguageimagepretraining}.
    
    \item \textbf{BLIP-2 Instruct}: Instruction-tuned version of BLIP-2, optimized for multimodal question-answering and instruction-following tasks.
    
    \item \textbf{Gemini 1.5 Flash}: A state-of-the-art multimodal large language model by Google, evaluated via its vision-text generation interface.

\end{itemize}

All models are evaluated in inference mode using the same set of radiology images to compare domain alignment and caption quality. 

\subsection{Fine-tuning Setup}
BLIP is fine-tuned using a learning rate of $5 \times 10^{-5}$ with the AdamW optimizer and gradient accumulation over 4 steps. We use mixed precision training via PyTorch’s \texttt{GradScaler}. The model is trained for 1–3 epochs with early stopping based on validation loss. Unless otherwise specified, we fine-tune the full model, including both the vision encoder (`ViT`) and the text decoder (`BERT`). All models are trained and evaluated using a single NVIDIA A100 GPU. We use the Hugging Face \texttt{transformers} and \texttt{datasets} libraries for training, checkpointing, and evaluation.

\subsection{Attention Interpretability}
To interpret the model's predictions, we extract decoder cross-attention weights from the final layer of BLIP’s BERT decoder during inference. Forward hooks are registered on the decoder blocks to capture attention tensors, which are averaged across attention heads and visualized as heatmaps.

The attention maps are resized to match the original image dimensions and overlaid using OpenCV-based color maps. For each generated token, we render a corresponding heatmap, enabling qualitative analysis of which image regions influenced specific word predictions. This visualization provides insights into visual grounding and helps identify potential failure modes or attention misalignment.

\subsection{Evaluation Metrics}
To evaluate the quality and relevance of generated captions, we use a combination of lexical, semantic, and vision-language alignment metrics:

\begin{itemize}
     \item \textbf{CIDEr:} Measure n-gram overlap between generated and reference captions. CIDEr emphasizes consensus and rewards rarer, content-rich n-grams.
    \item \textbf{SPICE:} Measures semantic similarity by comparing scene graph representations, and is useful for evaluating object-attribute relationships in captions.
    \item \textbf{BERTScore:} Computes similarity between generated and reference captions using contextual embeddings from pretrained language models. We report BERTScore using both the original BERT and the domain-specific Bio-ClinicalBERT model.
    \item \textbf{Cosine-Similarity:} Measures sentence-level semantic alignment between generated and reference captions using mean-pooled Bio-ClinicalBERT embeddings.
\end{itemize}

All text-based metrics are computed using the \texttt{evaluate} library from Hugging Face.For Bio\_ClinicalBERT-based metrics, we use the pretrained model from Hugging Face’s \texttt{emilyalsentzer/Bio\_ClinicalBERT}.

\subsection{Ablation Study}
To better understand the contribution of different components of the BLIP model, we conduct an ablation study using three configurations:
\begin{itemize}
  \item \textbf{Full Fine-Tuning}: Both the vision encoder and text decoder are updated during training.
  \item \textbf{Decoder-Only Fine-Tuning}: Only the BERT decoder is updated; the ViT encoder is frozen.
  \item \textbf{Encoder-Only Fine-Tuning}: Only the ViT encoder is trained; the decoder remains frozen.
\end{itemize}

This study helps isolate whether improvements (both performance and efficiency) are driven by language adaptation (decoder), visual domain alignment (encoder), or both. We report results separately in the Results section.

\section{Results}
\subsection{Experimental Setup}
All models were evaluated on a held-out validation subset of the ROCO dataset containing diverse radiology modalities (e.g.,  X-rays, MRI, CT, ultrasound, sonography). For fairness, all models generated captions using the same images and a beam search decoding strategy (beam size = 4, max length = 128).

\subsection{Quantitative Evaluation}
Table \ref{tab:metrics} reports the captioning performance of each model across four standard metrics: CIDEr, BLEU-4, METEOR, and CLIP similarity. The fine-tuned BLIP model consistently outperformed the base BLIP and other baselines, confirming that domain-specific adaptation improves some level of alignment with reference captions. However, these metrics do not fully reflect the clinical reliability of the generated outputs.

\begin{table*}[h]
\centering
\caption{Evaluation metrics for all models. BERTScore is reported using both Bio\_ClinicalBERT and Original BERT.}
\label{tab:metrics}
\scriptsize
\begin{tabular}{l|c|ccc|ccc|c|c}
\toprule
\textbf{Model} & 
\textbf{CosSim} & 
\multicolumn{3}{c|}{\textbf{BERTScore (Bio\_ClinicalBERT)}} & 
\multicolumn{3}{c|}{\textbf{BERTScore (Original BERT)}} & 
\textbf{CIDEr} & 
\textbf{SPICE} \\
\cline{3-5} \cline{6-8}
 & & \textbf{P} & \textbf{R} & \textbf{F1} & \textbf{P} & \textbf{R} & \textbf{F1} & & \\
\midrule
BLIP Base & 
0.8498 & 
0.7438 & 0.6764 & 0.7078 & 
0.5054 & 0.4122 & 0.4518 & 
0.0294 & 0.0171 \\
BLIP Fine-Tuned & 
0.8943 & 
0.7475 & 0.7118 & \textbf{0.7284} & 
\textbf{0.5539} & 0.4984 & \textbf{0.5221} & 
\textbf{0.0917} & \textbf{0.0409} \\
BLIP-2 & 
0.8663 & 
\textbf{0.7487} & 0.6840 & 0.7142 & 
0.5117 & 0.4174 & 0.4577 & 
0.0499 & 0.0315 \\
BLIP-2 Instruct & 
0.6777 & 
0.7502 & 0.6412 & 0.6908 & 
0.4065 & 0.2709 & 0.3224 & 
0.0051 & 0.0045 \\
Gemini 1.5 Flash & 
\textbf{0.9084} & 
0.6784 & \textbf{0.7375} & 0.7061 & 
0.4495 & \textbf{0.5580} & 0.4953 & 
0.0 & 0.0 \\  
ViT + GPT2 & 
0.8051 & 
0.6794 & 0.6368 & 0.6566 & 
0.3863 & 0.3259 & 0.3520 & 
0.0052 & 0.0048 \\
\bottomrule
\end{tabular}
\end{table*}

\begin{table*}[h]
\centering
\caption{Clinical evaluation of BLIP Base and Fine-Tuned captions compared to ground truth across four medical image cases.}
\label{tab:combined_diagnostic_evaluation}
\scriptsize
\resizebox{\textwidth}{!}{%
\begin{tabular}{|p{3.2cm}|p{2.6cm}|l|c|c|c|c|c|}
\hline
\textbf{Image (Type)} & \textbf{Model} & \textbf{Modality} & \textbf{Laterality} & \textbf{Correct Finding} & \textbf{Anatomical Specificity} & \textbf{Verdict} \\
\hline

\multirow{3}{*}{\shortstack{Chest X-ray\\(CWP)}} 
  & GT           & Radiograph (Frontal) & Bilateral & Reticulonodular + fibrosis + nodules & Apical lung zones & Ground truth \\
  & BLIP     & \ding{51} Radiograph & \ding{51} Bilateral & \ding{55} Pleural effusion (not GT) & \ding{55} No fibrosis/nodules & \ding{55} Incorrect \\
  & Fine-Tuned & \ding{51} Radiograph & \ding{51} Bilateral & \ding{55} Pleural effusion (hallucinated) & \ding{55} Missed disease pattern & \ding{55} Incorrect \\
\hline

\multirow{3}{*}{\shortstack{Brain MRI\\(Splenium)}} 
  & GT           & Axial T2 MRI & Midline (Symmetric) & Hyperintensity & Splenium of corpus callosum & Ground truth \\
  & BLIP     & \ding{55} Generic & \ding{55} Not Specified & \ding{55} None & \ding{55} None & \ding{55} Poor \\
  & Fine-Tuned & \ding{51} MRI & \ding{55} Left only & \ding{51} Hyperintensity (generic) & \ding{55} Wrong location & \ding{55} Misleading \\
\hline

\multirow{3}{*}{\shortstack{Knee X-ray\\(Avulsion)}} 
  & GT           & Radiograph (Lateral) & Right & Avulsion fracture & Patella (lower pole) & Ground truth \\
  & BLIP     & \ding{55} Generic & \ding{55} Left & \ding{51} Fracture (generic) & \ding{55} No detail & \ding{55} Poor \\
  & Fine-Tuned & \ding{51} Radiograph & \ding{51} Right & \ding{55} Osteolytic (wrong) & \ding{55} Missed patella/fracture & \ding{55} Misleading \\
\hline

\multirow{3}{*}{\shortstack{Abdominal US\\(Hydronephrosis)}}
  & GT           & Ultrasound & Right & Hydronephrosis with contents & Enlarged right kidney & Ground truth \\
  & BLIP    & \ding{55} Generic scan & \ding{55} Not Specified & \ding{55} None & \ding{55} No organ/pathology & \ding{55} Poor \\
  & Fine-Tuned & \ding{51} Ultrasound & \ding{51} Right & \ding{55} Missed hydronephrosis & \ding{51} Kidney identified & \ding{55} Incomplete \\
\hline

\end{tabular}
}
\end{table*}

\subsection{Qualitative Comparison}
Table \ref{tab:image_caption_grid} presents example captions from the base BLIP, and fine-tuned BLIP model for the same radiology images. While the base BLIP model produces generic and modality-agnostic descriptions (e.g., "a medical scan of the brain"), the fine-tuned model shows improved use of anatomical and modality-specific terminology (e.g., "MRI showing a hyperintense lesion").

Despite this improvement, the fine-tuned BLIP often introduces clinically incorrect details — such as incorrect laterality, hallucinated pathologies, or missed key findings — which can be harmful in medical applications. Our qualitative evaluation in table \ref{tab:combined_diagnostic_evaluation} highlights that improved metrics do not guarantee clinical correctness, emphasizing the need for medically-aware architectures or post-hoc validation in high-stakes settings.





\begin{table*}[htbp]
\centering
\caption{Comparison of BLIP and Fine-Tuned BLIP captions against Ground Truth (GT) across multiple medical images.}
\label{tab:image_caption_grid}
\scriptsize
\renewcommand{\arraystretch}{1.2}
\begin{tabular}{|c|p{4.2cm}|p{4.2cm}|p{4.2cm}|}
\hline
\textbf{Image Type} & \textbf{BLIP Caption} & \textbf{BLIP Fine-Tuned Caption} & \textbf{Ground Truth (GT)} \\
\hline
\includegraphics[width=2.8cm,height=2.8cm]{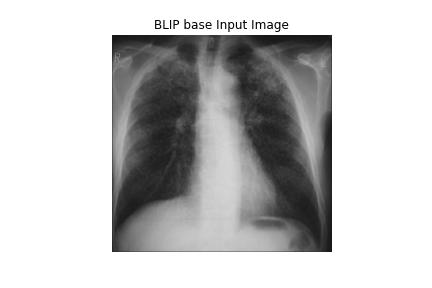} &
a medical scan of large chest &
chest x-ray showing bilateral pleural effusion &
reticulonodular shadowing with apical fibrosis and nodules (CWP) \\
\hline

\includegraphics[width=2.8cm,height=2.8cm]{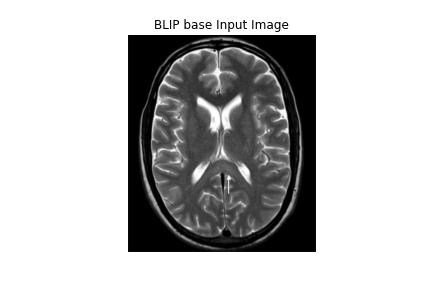} &
a medical scan of the brain &
hyperintense lesion in the left frontal lobe (arrow) &
symmetric hyperintensity in splenium of corpus callosum (arrow) \\
\hline

\includegraphics[width=2.8cm,height=2.8cm]{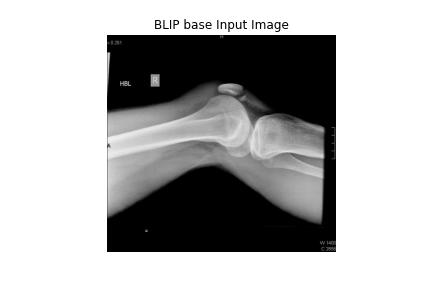} &
fracture of the left knee &
radiograph of the right knee showing a large osteolytic lesion in the proximal &
avulsion fracture of the lower pole of the right patella \\
\hline

\includegraphics[width=2.8cm,height=2.8cm]{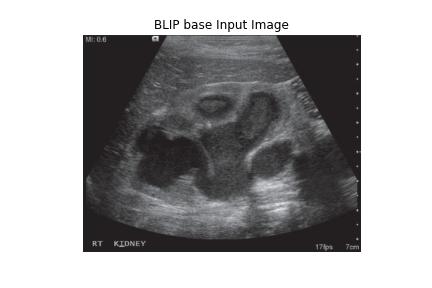} &
small, flat, black-and-white area &
ultrasound image of the right kidney &
hydronephrotic right kidney with thick material inside \\
\hline
\end{tabular}
\end{table*}

\begin{figure*}[t]
\centering
\makebox[\textwidth][c]{%
\begin{minipage}[t]{0.18\textwidth}
  \centering
  \includegraphics[width=\linewidth, height=3.0cm, keepaspectratio]{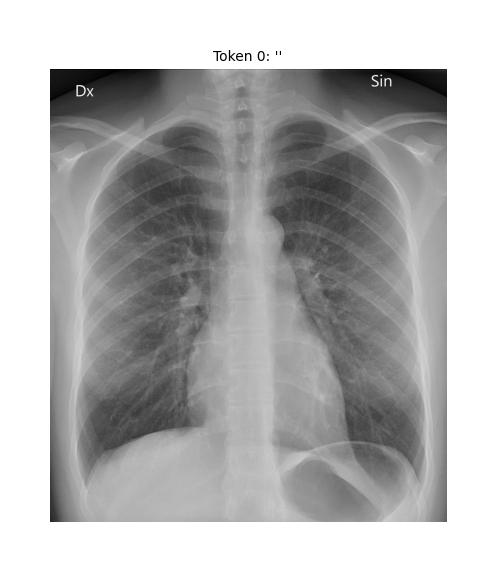}
  \vspace{2pt}
  \scriptsize
  \textbf{Ground Truth} \\[2pt]
  Caption: Cardiomegaly with bilateral infiltrates.
\end{minipage}
\hfill
\begin{minipage}[t]{0.18\textwidth}
  \centering
  \includegraphics[width=\linewidth, height=3.5cm, keepaspectratio]{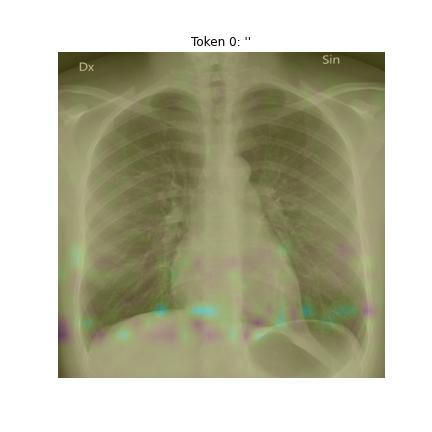}
  \vspace{2pt}
  \scriptsize
  \textbf{BLIP (Base)} \\[2pt]
  Caption: a chest x - ray with a large, open chest.
\end{minipage}
\hfill
\begin{minipage}[t]{0.18\textwidth}
  \centering
  \includegraphics[width=\linewidth, height=3.5cm, keepaspectratio]{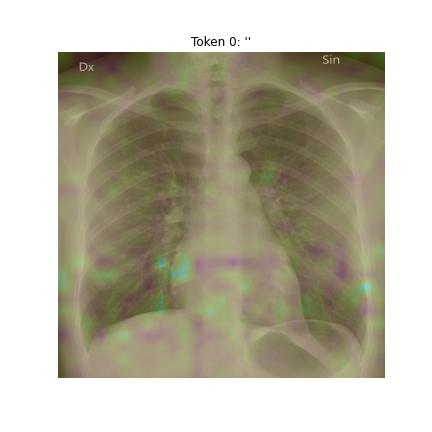}
  \vspace{2pt}
  \scriptsize
  \textbf{BLIP Fine-tuned (Full)} \\[2pt]
  Caption: chest x - ray showing bilateral pleural effusion.
\end{minipage}
\hfill
\begin{minipage}[t]{0.18\textwidth}
  \centering
  \includegraphics[width=\linewidth, height=3.5cm, keepaspectratio]{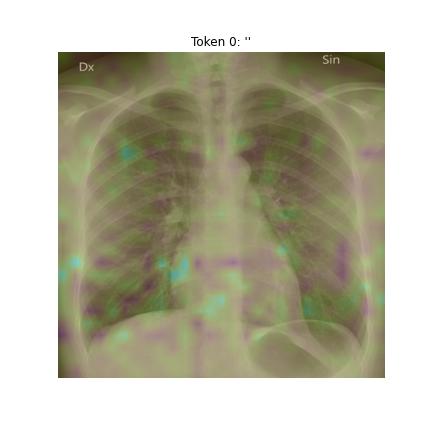}
  \vspace{2pt}
  \scriptsize
  \textbf{BLIP Fine-tuned (Decoder Only)} \\[2pt]
  Caption: chest x - ray showing a large right - sided pleural effusion.
\end{minipage}
\hfill
\begin{minipage}[t]{0.18\textwidth}
  \centering
  \includegraphics[width=\linewidth, height=3.5cm, keepaspectratio]{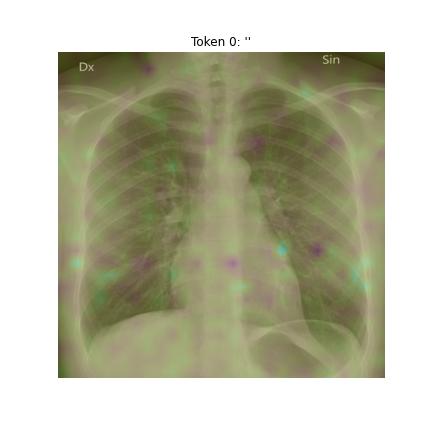}
  \vspace{2pt}
  \scriptsize
  \textbf{BLIP Fine-tuned (Encoder Only)} \\[2pt]
  Caption: chest x - ray showing bilateral pleural effusion.
\end{minipage}
}

\caption{Attention visualization across models for the same chest X-ray. Each method generates different interpretations, showing how model fine-tuning affects focus and caption quality.}
\label{fig:attn}
\end{figure*}

\subsection{Attention Visualization}
To assess whether the model’s generated tokens align with relevant image regions, we visualize cross-attention maps for multiple BLIP variants on the same chest X-ray image (Figure~\ref{fig:attn}). The base BLIP model distributes attention diffusely and often fixates on irrelevant areas, producing a generic and clinically inaccurate caption. In contrast, fine-tuned models (both encoder-only and full fine-tuning) exhibit more localized attention over key anatomical regions—such as the lungs—when generating terms like “pleural effusion.”

Interestingly, while fine-tuned models attend more precisely, they can still hallucinate findings not present in the ground truth. For instance, some versions focus on the correct regions yet describe unilateral effusions or overstate severity. This indicates that improved spatial grounding does not always ensure clinical correctness. As such, attention maps are helpful for interpretability but must be paired with clinical verification when deploying vision-language models in medical settings.

\subsection{Ablation Study}
Table~\ref{tab:ablation} presents results from three BLIP configurations: full fine-tuning, decoder-only, and encoder-only. Decoder-only fine-tuning performs competitively while being more resource-efficient. Encoder-only fine-tuning results in poorer performance, likely due to a lack of domain adaptation in the captioning logic.

While full fine-tuning achieves the highest scores across all metrics, decoder-only fine-tuning performs competitively and requires less training time. In our setup, encoder-only training completed an epoch in approximately 4 hours and 59 minutes, compared to 5 hours and 15 minutes for full fine-tuning—a modest but meaningful computational saving, especially in low-resource settings.

\begin{table*}[h]
\centering
\caption{Evaluation metrics for different fine-tuning strategies. BERTScore is reported using both Bio\_ClinicalBERT and Original BERT.}
\label{tab:ablation}
\scriptsize
\begin{tabular}{l|c|ccc|ccc|c|c}
\toprule
\textbf{Strategy} & 
\textbf{CosSim} & 
\multicolumn{3}{c|}{\textbf{BERTScore (Bio\_ClinicalBERT)}} & 
\multicolumn{3}{c|}{\textbf{BERTScore (Original BERT)}} & 
\textbf{CIDEr} & 
\textbf{SPICE} \\
\cline{3-5} \cline{6-8}
 & & \textbf{P} & \textbf{R} & \textbf{F1} & \textbf{P} & \textbf{R} & \textbf{F1} & & \\
\midrule
Encoder-Only & 
0.8925 & 
\textbf{0.7501} & 0.7103 & \textbf{0.7289} & 
\textbf{0.5558} & 0.4932 & 0.5200 & 
0.0822 & 0.0356 \\
Decoder-Only & 
0.8863 & 
0.7433 & 0.7049 & 0.7228 & 
0.5478 & 0.4873 & 0.5132 & 
0.0664 & 0.0275 \\
Full Fine-Tune & 
\textbf{0.8943} & 
0.7475 & \textbf{0.7118} & 0.7284 & 
0.5539 & \textbf{0.4984} & \textbf{0.5221} & 
\textbf{0.0917} & \textbf{0.0409} \\

\bottomrule
\end{tabular}
\end{table*}

\section{Conclusion}
In this paper, we explored the effectiveness of fine-tuning VLMs for medical image captioning using the ROCO dataset. We demonstrated that the pretrained BLIP model, while effective in general domains, generates generic and often clinically irrelevant captions when applied to radiology images without adaptation. Fine-tuning BLIP on domain-specific data significantly improved both lexical and semantic alignment, as shown through quantitative metrics and qualitative analysis of generated captions and attention maps.

Our ablation study showed that encoder-only fine-tuning offers a favorable performance-cost trade-off, while full fine-tuning yields the best results when resources permit. However, qualitative analysis revealed that even fine-tuned models may hallucinate, misidentify anatomy, or omit key findings—highlighting limitations of standard metrics and the need for clinically grounded evaluation.

We also explored parameter-efficient fine-tuning (LoRA) and multimodal captioning via LLaVA-Med. Although not fully implemented due to technical constraints, these directions informed our approach and suggest promising areas for future work. Moving forward, incorporating structured medical knowledge and safety constraints will be essential for reliable deployment in clinical settings.

{\small
\bibliographystyle{ieee_fullname}
\bibliography{egbib}
}
\end{document}